\documentclass{edp-conf}
\usepackage{graphicx}
\def\beqa{\begin{eqnarray}}
\def\eeqa{\end{eqnarray}}
\def\beq{\begin{equation}}
\def\eeq{\end{equation}}
\def\acknowledgements{\par\addvspace{17pt}\footnotesize\rmfamily\trivlist\item[]
}
\def\endacknowledgements{\endtrivlist\addvspace{6pt}}
\begin{document}

\TitreGlobal{SF2A 2004}

\title{Quasi-equilibrium sequences of binary strange quark stars in 
general relativity}
\author{Francois Limousin}\address{Laboratoire de l'Univers et de ses 
Th\'eories,UMR 8102 du C.N.R.S., Observatoire de Paris, Universite Paris 7, F-92195 Meudon Cedex, France}
\author{Dorota Gondek-Rosi\'nska $^1$}\address{also at Nicolaus
  Copernicus Astronomical Center, Bartycka 18, 00-716 Warszawa, Poland  and Institute of Astronomy, University of Zielona Gora, Lubuska 2, 65-265, Zielona Gora
, Poland
}
\author{Eric Gourgoulhon $^1$}
\runningtitle{F. Limousin, D.Gondek-Rosinska and E.Gourgoulhon : Strange stars }
\setcounter{page}{237}
\index{Francois Limousin, A.}
\index{Dorota Gondek-Rosi\'nska, B.}
\index{Eric Gourgoulhon, C.}

\maketitle
\begin{abstract}

Inspiraling compact binaries are expected to be the strongest sources
of gravitational waves for VIRGO, LIGO and other laser
interferometers. We present the first computations of
quasi-equilibrium sequences of compact binaries containing two strange
quark stars (which are currently considered as a possible alternative
to neutron stars). We study a precoalescing stage in the conformal
flatness approximation of general relativity using a multidomain
spectral method. A hydrodynamical treatment is performed under the
assumption that the flow is irrotational. 

\end{abstract}

%
\section{Introduction}

One of the most important prediction of Einstein theory of relativity
is gravitational radiation. Since the very precise measurement of the
orbital decay in the binary pulsar B1913+16 system by Hulse and
Taylor, the existence of gravitational waves (GW) has been indirectly
proved and general relativity has passed another quite constraining
test. Due to the emission of GW, binary neutron
stars (NS) decrease their orbital separation and finally merge. The
evolution of a binary system can be separated into three phases :
point-like inspiral where orbital separation is much larger than the
NS radius, hydrodynamical inspiral where orbital separation is just a
few times larger than the radius of the NS so that hydrodynamics play
an important role, and merger in which the two stars coalesce
dynamically.

 The GW signal of the terminal phases (the hydrodynamical phase or the
 merger phase) of inspiraling binary can bring the information about
 the stellar structure. In particular it may be possible to impose
 constraints on the equation of state of NS. It is still an open
 question whether the core of NS consists mainly of superfluid
 neutrons or an exotic matter like kaon condensations, pion
 condensations or strange quark matter. As suggested by Bodmer (1971)
 the absolute true ground state of nuclear matter may be a state of
 deconfined up, down and strange quarks (since energy per baryon for
 strange matter is lower than the energy per baryon for Fe$^{56}$).
 If it is true then objects made of such matter so called strange
 stars (SS) could exists (Witten 1984). SS are currently considered as
 a possible alternative to NS as compact objects (see e.g.  Madsen
 1999 for a review and Gondek-Rosi\'nska et al. 2003).

Up to now, majority of the relativistic calculations of the terminal
phase of inspiral have been done for binary systems containing NS
described by a simplified equation of state of dense matter so called
polytropic EOS (Taniguchi \& Gourgoulhon 2003).  In
the paper we present the results of our studies on the hydrodynamical
phase of inspiraling binary systems containing equal mass strange
stars (according to the recent population synthesis calculations
(Bulik, Gondek-Rosi\'nska and Belczy\'nski, 2004) a significant
fraction of the observed binary NS in GW will contain stars
with masses $\sim 1.4\ M_{\odot}$).  We compare the evolution of SS-SS
systems with NS-NS systems in order to find any characteristic
features in the GW waveform that will help to distinguish between SS
and NS.

\section{Strange quark stars and stellar models} 

As already mentioned, SS are composed of deconfined up,
down and strange quarks. Typically, they are modeled with an equation
of state based on the MIT-bag model in which quark confinement is
described by an energy term proportional to the volume (Fahri et al.,
1984). SS are self-bound objects, having high density ($> 10^{14}
{\rm g\ cm}^3 $) at the surface.  There are three
physical quantities entering the MIT-bag model: the mass of the strange
quarks, $m_{\rm s}$, the bag constant, $B$, and the strength of the
QCD coupling constant $\alpha$.   
In the numerical calculations reported in the present paper
we consider three different MIT-bag models  (corresponding to three
different sets of the model parameters):

{\bf SQS0} - the standard MIT bag model: $m_{\rm s}c^2=200\ {\rm MeV}$,
$\alpha=0.2$, $B=56~{\rm MeV/fm^3}$;
{\bf SQS1} -  the simplified MIT bag model with  $m_{\rm s}=0$, $\alpha=0$;
$B=60\ {\rm MeV/fm^3}$;
{\bf SQS2} - the ''extreme'' MIT bag model (relatively low strange 
quark mass and $B$ but high $\alpha$) : $m_{\rm s}c^2=100\ {\rm MeV}$,
$\alpha=0.6$, $B=40\ {\rm MeV/fm^3}$

\section{Basic assumptions}

In the hydrodynamical phase, since the timescale of orbital shrinking
due to the emission of GW is longer than the orbital period, one may
consider a binary NS system to be in quasiequilibrium state. For given
EOS, we construct so called {\em evolutionary sequence} by calculating
a sequence of quasiequilibrium configurations with constant baryon
mass for decreasing orbital separation. The second assumption is to
consider a conformally flat metric, which corresponds to the {\em
Isenberg-Wilson-Mathews} approximation of general relativity. In this
approximation, the spacetime metric takes the form: \beq ds^2 =-(N^2
-B_i B^i) dt^2 -2B_i \, dt \, dx^i +A^2 f_{ij} dx^i dx^j,
  \label{eq:metric}
\eeq 
where $N$ is the lapse, $A$ the conformal factor, $B^i$ the shift vector
and $f_{ij}$ the flat spatial metric.
Another assumption concerns the fluid motion inside
each star, here we considered irrotational binaries.

\section{Results and conclusions}


\begin{figure}[h]
  \centering
  \includegraphics[angle=0, width=7cm]{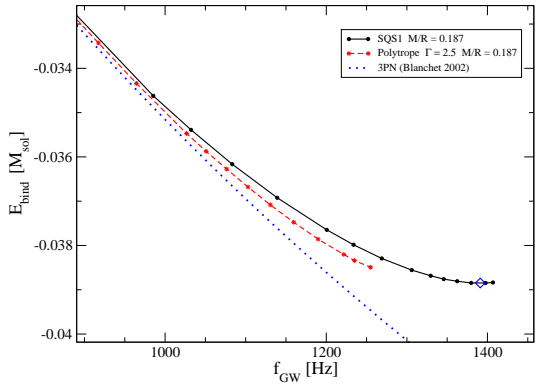}
  \caption{Orbital binding energy of an equal mass binary system
    ($M_1=M_2=1.35\ M_\odot$) versus frequency of GW along two
    irrotational equilibrium sequences. Solid and long-dashed lines
    correspond to strange quark stars described by MIT bag model and
    NS described by polytropic EOS respectively. A diamond indicates
    the marginally stable orbit.  A dotted line corresponds to 3rd
    post-Newtonian calculations for point masses. }
  \label{Ebin_cor}
\end{figure}

In order to calculate the last orbits of inspiral phase of binary NS
and SS we use highly accurate numerical code which solves the five
elliptic equations for the gravitational field (for $A$, $N$ and
$B^i$), supplemented by an elliptic equation for the velocity
potential in the case of irrotational flows (see Limousin,
Gondek-Rosi\'nska \& Gourgoulhon 2004 for boundary conditions in the
case of SS).  In Fig. 1 we show the evolution of equal mass binary NS
(a long-dashed line) and SS described by SQS1 model (a solid line) having
total gravitational mass $2.7 M_\odot$ at infinity. The binding energy
is defined as the difference between $M_{ADM}$ (see Taniguchi \&
Gourgoulhon 2003) and the total mass of the system at infinity. 
We see a good agreement with 3PN calculations for big distances i.e. small
frequencies, where the internal structure of the stars is not the
predominant effect.

 In order to compare results for SS and NS we take a polytrope giving
the same gravitational mass and the same compaction parameter $GM/Rc^2$ at
infinity as obtained for static SS.  The minimum of energy (shown as a
diamond) for the evolutionary sequence of SS corresponds to the
appearance of a dynamical instability for binaries (the innermost
stable circular orbit (ISCO)). The frequency of the ISCO is a
potentially observable parameter by the GW detectors. We don't see the
ISCO for NS. The sequence of NS terminates by the mass-shedding limit
(corresponding to exchange of matter between two stars). Different
evolutions of NS and SS stem from the fact that SS are principally bound by an
additional force, strong interaction between quarks (for the same
distances there are less deformed than NS).

\begin{figure}[h]
  \centering
  \includegraphics[angle=0, width=7cm]{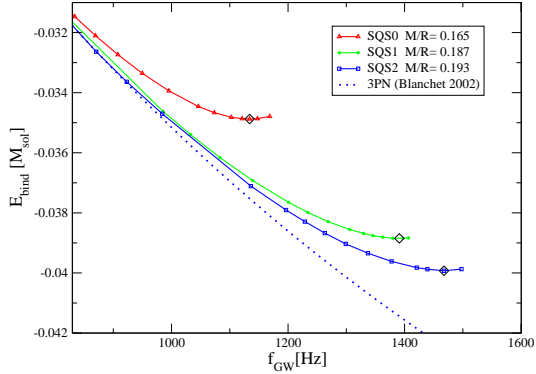}
  \caption{Orbital binding energy  versus frequency of GW 
    along evolutionary sequences of irrotational strange quark star
    binaries described by three different MIT bag models. }
  \label{Ebin_irr}
\end{figure}
In Fig 2. we present the evolution of binary SS, with total mass 2.7
$M_\odot$, described by three different sets of EOS parameters of the
MIT bag model. We see that the frequency of GW at the ISCO strongly
depends on the compaction parameter - the higher it is, the
higher the frequency at the ISCO. Detection of GW may help to impose
constraints on the EOS of NS and SS and to find the ground state of
matter at high densities.

\acknowledgements 
Partially supported by the KBN grants 5P03D.017.21
and PBZ-KBN-054/P03/2001; by the ``Bourses de recherche 2004 de la
Ville de Paris'' and by the Associated European Laboratory Astro-PF
(Astrophysics Poland-France).  
\endacknowledgements

\end{document}